\documentclass[pre,twocolumn,groupedaddress,showpacs,pdfoutput=1]{revtex4-1}
\usepackage{graphicx}
\usepackage{epsfig}
\usepackage{amssymb,amsmath,amsfonts,hyperref}
\usepackage{wasysym}
\usepackage{bm}
\usepackage{latexsym}
\usepackage{eucal}
\usepackage{verbatim}
\usepackage[normalem]{ulem}
\usepackage{color}

\newcommand{\ee}{\end{equation}}

\newcommand{\eea}{\end{eqnarray}}

\begin{document}

\title{Numerical Investigations of SO(4) Emergent Extended Symmetry in\\
Spin-1/2 Heisenberg Antiferromagnetic Chains}
\author{Pranay Patil}
\author{Emanuel Katz}
\author{Anders W. Sandvik}

\affiliation{Department of Physics, Boston University, 590 Commonwealth Avenue, Boston, Massachusetts 02215, USA}

\begin{abstract}
The antiferromagnetic Heisenberg chain is expected to have an 
extended symmetry, [SU(2)$\times$SU(2)]/Z$_2$, in the infrared limit, 
whose physical interpretation is that the spin and dimer order 
parameters form the components of a common 4-dimensional pseudovector. Here we 
numerically investigate this emergent symmetry using quantum Monte Carlo 
simulations of a modified Heisenberg chain (the J-Q model) in which the 
logarithmic scaling corrections of the conventional Heisenberg chain can 
be avoided. We show how the two- and three-point spin and dimer correlation 
functions approach their forms constrained by conformal field theory 
as the system size increases and numerically confirm the expected
effects of the extended symmetry on various correlation functions.
\end{abstract}

\maketitle

\section{Introduction}
The development of the theory of deconfined quantum criticality
\cite{Senthil} has reignited interest in critical quantum systems which show an 
extended symmetry in the thermodynamic limit \cite{Nahum,NahumX,
qin2017duality}. 
An example of the same can be seen in the spin-1/2 
Heisenberg antiferromagnetic chain, where it is expected \cite{A,HA}
that the microscopic SU(2) symmetry extends to an [SU(2)$\times$SU(2)]/Z$_2$ 
symmetry. The ground state of this Hamiltonian is known to be critical and has 
been shown to have scale invariant behavior analytically \cite{GiamS} and 
numerically\cite{SandvikHS}. In this work, the emergence of the extended
symmetry is connected to the behavior of two- and three-point correlation
functions thus providing a bridge between the continuum field theory and
lattice correlation functions.

Here we first connect the emergent symmetry to lattice correlation functions,
then numerically study these and show that they reflect the emergent symmetry. The 
correlation functions have strict functional forms which are controlled by the 2D 
conformal field theory which describes the emergent physics. The emergent [SU(2)$\times$SU(2)]/Z$_2$ 
symmetry is manifested in the three components of the N\'eel order parameter and the dimer order 
parameter (which quantifies the spin-Peierls order) forming an SO(4) symmetric pseudovector\cite{Tsvelik}.
The Z$_2$ reduction to the SU(2)$\times$SU(2) is required as it is a ``double
cover" of SO(4)\cite{LieBook}, i.e, there are two sets of SU(2) matrices which
generate the same SO(4) rotation.
The three-point functions of these order parameters can yield useful information about the emergent symmetry but must be treated with care, as a naive addition 
of scaling dimensions to infer the exponent of the power-law decay fails. An example of this will 
be presented here with the spin-spin-dimer three point function in the Heisenberg chain. We
discuss how the connections between the continuum and lattice version of the correlation 
functions have to be carefully considered in order to predict the correct power-law decay
of the three-point function.

The outline of the paper is as follows:
In Sec.~\ref{sec2} we review the predictions from Renormalization Group (RG)
analysis and examine how the extended symmetry affects the correlation
functions of the continuum versions of the lattice spin and dimer operators.
In Sec.~\ref{sec3}, we present the manifestation of a CFT description on the
correlation functions and benchmark these findings against numerical simulations
of the Transverse Field Ising Model (TFIM) on a periodic chain.
We stitch together the results of these two sections in Sec.~\ref{sec4}
and predict the complete functional forms for the correlation functions
and then present numerical evidence to support the same. We briefly summarize 
the study and discuss possible future applications of numerical CFT tests
in Sec.~\ref{sec5}.

\section{\label{sec2}Emergence of [SU(2)$\times$SU(2)]/Z$_2$ Symmetry}

The spin-1/2 Heisenberg chain, with the Hamiltonian
\begin{equation}\label{HHam}
H=\sum_{i=1}^N \mathbb{\vec{S}}_i\cdot\mathbb{\vec{S}}_{i+1}
\end{equation}
can be transformed into a system of interacting spinless fermions 
of two species using the transformation 
$\vec{S}_n = \frac{1}{2}{\psi^{\dagger}}_n^i\vec{\bm{\sigma}}^j_i\psi_{nj}$ 
\cite{A,HA}, where $\psi_n^i$ is a spin doublet and repeated indices imply
summation over the range of values that the index can take (as we will use 
thoughout this work). To take the continuum limit,
we will reiterate the series of arguments presented in \cite{A,HA}
and will use this process to define some quantities that we will use later.
Each fermion in the doublet can be rewritten as two new fermions,
\begin{equation}
\psi_n^j\simeq [i^n\psi^j_L(n\pm\frac{1}{2})+(-i)^n\psi^j_R(n\pm\frac{1}{2})],
\end{equation}
(plus and minus for even and odd $n$, respectively) which is an exact
transformation up to an overall factor. This is motivated by the expectation
that the free-fermion ground state would have all states with
$|k|< \pi/2a$ occupied and thus only Fourier modes with
$k\simeq \pm\pi/2a$ would be important\cite{H1980}. 
Thus, we understand the left (L) and 
right (R) fermion operators to be ``locally" constant and to be slowly varying
at the scale of lattice separation. These will ultimately form the operators
of the continuum field theory. We can now write the spin operator on the
lattice, using current operators 
($\vec{J}_L = {\psi^{\dagger}}_L^i\vec{\bm{\sigma}}^j_i\psi_{Lj}, 
\vec{J}_R = {\psi^{\dagger}}_R^i\vec{\bm{\sigma}}^j_i\psi_{Rj}
$) and a fermion biliear 
(${\bm{G}}^i_j = {\psi^{\dagger}}_L^i\psi_{Rj}$) as
\begin{equation}\label{Latt1}
\mathbb{S}^i_n = a(J_L^i+J_R^i)+(-1)^na\mathrm{Tr}[(\bm{G}-\bm{G}^{\dagger})\bm{\sigma}^i].
\end{equation}
Here we have used script letters for lattice operators 
and bold font for matrices 
and we will continue to maintain these conventions throughout this text.
The operators $\vec{J}_L,\vec{J}_R,\bm{G}$ are defined at the same lattice
position $n$ as the spin operator but this is not explicitly indicated
to keep the equations unencumbered. This form of the spin operator
can be substituted into the Hamiltonian of Eqn~\eqref{HHam}, which upon
coarse graining has the following continuum limit,
\begin{equation}
H= (a/2) \int dx [\vec{J}_L\cdot\vec{J}_L+\vec{J}_R\cdot\vec{J}_R+2\vec{J}_L\cdot\vec{J}_R] + ...,
\end{equation}
where $a$ is the lattice spacing\cite{A}.

Note also that at this stage we have only one SU(2) symmetry which comes
along with the 3D rotation symmetry that the microscopic model has.
This is manifest in each of $\vec{J}_{L/R}$ but they are not free to
turn through different arbitrary angles due to the $\vec{J}_L\cdot\vec{J}_R$
term which keeps the relative angle between them fixed.
Assuming that this Hamiltonian flows to the free fermion fixed point,
which has the Hamiltonian
\begin{equation}\label{Hfix}
H_{fixed}=  \int dx [\vec{J}_L\cdot\vec{J}_L+\vec{J}_R\cdot\vec{J}_R],
\end{equation}
it can be shown that the term that couples left and right currents
is irrelavant under RG flow \cite{A,HA} for this particular fixed point. 
This is not true for all perturbations to the fixed point Hamiltonian
and thus was checked explicitly \cite{A,HA} for the $\vec{J}_L\cdot\vec{J}_R$
term. Thus we see that this line
of reasoning leads us to believe that in the thermodynamic limit we should
be left with the free fermion fixed point, which is also described by the 
$k=1$ Wess-Zumino-Witten (WZW) conformal field theory \cite{241}.  

To understand how this decoupling of the currents affects correlation
functions of spin and dimer operators, we must first connect the primary
operators of the CFT to these order parameters. Once we have done this,
we can use the constraints that the extended symmetry places on the
correlation functions of the primary operators to understand the 
correlations of the measurable orders.

The primary operators of the $k=1$ WZW theory that we are going to be
interested in are $[\bm{J_L},\bm{J_R},\bm{g}]$, which are the left and
right currents and the primary field with scaling dimension $1/2$.
These are all SU(2) matrices, although the currents form matrices which
belong to the Hermitian subset of SU(2), which are described by SO(3) vectors.
This can be seen by observing that $\vec{J}_{L/R}$ in the fixed point
Hamiltonian [Eq.~\eqref{Hfix}] are SO(3) vectors and thus the matrices
to represent these must be written as
\begin{equation}
\bm{J_{L/R}}=J_{L/R}^a \bm{\sigma^a},
\end{equation}
where $J_{L/R}^a$ form the components of $\vec{J}_{L/R}$.
This structure is also justified by the framework of the 2D CFT, which
requires independent generators of translations for $z$ and $\bar{z}$
(conjugate variables in the complex plane). In the Virasoro algebra
of the 2D CFT \cite{Fran}, these would usually be called $J(z)$ and 
$\bar{J}(\bar{z})$ and in the case of the left(right) fermion,
$z=x+it$ ($\bar{z}=x-it$) would encode its space-time position.

The primary field $\bm{g}$ is made out of the continuum versions of the
lattice operators which we shall denote as $(S^a,D)$. The components $S^a$
form the continuum spin operators and $D$ represents the continuum dimer
operator. These together form an SO(4) pseudovector which is 
embedded in $\bm{g}$ through
\begin{equation}
\bm{g}=S^a\ i\bm{\sigma^a}+D\bm{I},
\end{equation}
as any general SU(2) matrix can be expanded in this manner. The continuum
versions of spin and dimer will be mapped back to the lattice variables in
the next section. 

As mentioned earlier, the left and right currents can turn through different
arbitrary angles at the fixed point and these SO(3) rotations can be written
in terms of transformations on the SU(2) matrices as
\begin{equation}\label{LJL}
\bm{{J}_L}=J_{L}^a \bm{\sigma^a}\to\bm{L{J}_LL^{\dagger}}=J_{L}^{'a} \bm{\sigma^a},
\end{equation}
\begin{equation}\label{RJR}
\bm{{J}_R}=J_{R}^a \bm{\sigma^a}\to\bm{R{J}_RR^{\dagger}}=J_{R}^{'a} \bm{\sigma^a},
\end{equation}
where $L$ and $R$ are the SU(2) rotation matrices. It is important to note 
here that these rotations do not mix left and right currents and keep the
2D conformal structure intact. The field $\bm{g}$ depends on $z$ and $\bar{z}$
by construction\cite{Fran} and thus is affected by both left and right
rotations. These rotations are reflected in $(S^a,D)$ through
\begin{equation}\label{LgR}
\bm{g}=S^a\ i\bm{\sigma^a}+D\bm{I}\to\bm{LgR^{\dagger}}=S^{'a}\ i\bm{\sigma^a}
+D^{'}\bm{I},
\end{equation}
which creates the new set $(S^{'a},D^{'})$. The matrices 
$(\bm{g},\bm{J_{L}},\bm{J_{R}})$ live on the complex plane formed by
space-time and so do their components. The correlation functions of
these components (which are the continuum spin, dimer, and current operators)
on the complex plane are of interest to us as they tell us what to expect
for the correlation functions of the lattice operators, which we will
investigate numerically later. We would also like to point
out here that all the correlation functions that we consider
in this text are connected correlation functions as
they are the ones which the CFT predicts. From this
point on, we will not explicitly mention that we are only
considering connected correlation functions.
For the continuum operators, the connected correlation functions we consider 
are the same as the naive correlation functions as all the operators have
a zero single body expectation (enforced by the CFT) 
value and this implies nothing needs
to be subtracted from the naive correlation function to get the connected
one. 

To extract the correlation functions
of $(S^a,D)$, we look at
\begin{equation}
\langle \mathrm{Tr}[\bm{{g}_{z_1}}\bm{{g^{\dagger}}_{z_2}}]\rangle =
\langle S^aS^a\rangle+\langle DD\rangle
\end{equation}
and see that the right hand side is non-zero as the arbitrary transformations
$\bm{L}$ and $\bm{R}$ leave the two point function of $\bm{g}$, as defined
here, unchanged through Eq.~\eqref{LgR} due to the cyclicity of the trace
and as $\bm{R^{\dagger}}\bm{R}=\bm{L^{\dagger}}\bm{L}=I$. 
Similarly, if we look at the transformation of the three point function,
\begin{eqnarray}
&&\langle \mathrm{Tr}[\bm{{g}_{z_1}}\bm{{g^{\dagger}}_{z_2}}\bm{{g}_{z_3}}]\rangle\to \\
&&~~~~~\langle \mathrm{Tr}[\bm{L{g}_{z_1}R^{\dagger}} 
\bm{R{g^{\dagger}}_{z_2}L^{\dagger}}
\bm{L{g}_{z_3}R^{\dagger}}]\rangle, \nonumber
\end{eqnarray}
we see that the only way to keep this invariant under arbitrary $\bm{L}$ and 
$\bm{R}$ would be to have this vanish. The vanishing of the three point
function then implies that all three point functions of $(S^a,D)$ (which
could be either of $\langle S^aS^aD\rangle,\langle S^aDS^a\rangle,
\langle DS^aS^a\rangle,\langle DDD\rangle$) would vanish. This can be seen
by writing down all the possible three point functions
of $\bm{g}$ and $\bm{g^{\dagger}}$ and solving for the spin and dimer
correlation functions. 

When discussing the lattice spin and dimer correlation functions, we will also 
need the continuum versions of the correlation functions of the currents and operators. 
The expressions which are of interest to us and robust against arbitrary $\bm{L}$ 
and $\bm{R}$ transformations are
\begin{equation}
\langle \mathrm{Tr}[\bm{{J}}_{L}\bm{{J}}_{L}]\rangle \sim \langle J_L^aJ_L^a\rangle,
\end{equation}
\begin{equation}
\langle \mathrm{Tr}[\bm{{J}}_{L}\bm{g}\bm{g^{\dagger}}]\rangle \sim \langle J_L^aS^aD\rangle+\langle J_L^aDS^a\rangle,
\end{equation}
and permutations of the second equation with $\bm{{J}_{L}}$ in different
positions. These equations also apply for the right currents by just switching
all $L\to R$. The non-vanishing nature of these correlation functions can be
seen again using the cyclicity of the trace and transformation equations
~\eqref{LJL},~\eqref{RJR} and ~\eqref{LgR}.

\section{\label{sec3}CFT and Correlation functions}

Due to the mapping to massless fermions, the spin-1/2 Heisenberg chain
is a 2D CFT and we can use the constraints the 2D CFT puts on the correlation
functions to understand their precise forms. A 2D CFT is defined on the
complex plane and the constraints that conformal symmetry requires restricts
the two point and three point correlation functions of primary operators
to behave as \cite{Fran}
\begin{equation}\label{Corrs1}
 \langle O_iO_j\rangle \sim \frac{\delta_{ij}}{|z|^{\Lambda_i\Lambda_j}},\\
\end{equation}
\begin{equation}\label{Corrs2}
 \langle O_iO_jO_k\rangle \sim \frac{1}{|z_{ij}|^{\Delta_{ij}}|z_{jk}|^{\Delta_{jk}}|z_{ki}|^{\Delta_{ki}}},
\end{equation}
where $O_i$ are the primary operators of the CFT and $\Lambda_i$ are their
scaling dimensions and $\Delta_{ij}=\Lambda_i+\Lambda_j-\Lambda_k$.
The scaling dimension of a primary operator is made up of two numbers
$h_i$ and $\bar{h}_i$ which indicate the scaling in $z$ and $\bar{z}$
respectively. These correlation functions can be factorised into two
pieces, where one depends on $z$, and the other on $\bar{z}$ and the
$\Lambda_i$'s are replaced by $h_i(\bar{h}_i)$. This is useful when 
$h_i\neq\bar{h}_i$, which may be the case for current operators, which 
generate translations in either $z$ or $\bar{z}$. For operators with
$h_i=\bar{h}_i$, we do not mention each of them separately, but only
indicate the total scaling dimension $\Lambda_i=h_i+\bar{h}_i$.

As these relations are valid on the infinite complex plane and our simulations
are done on a periodic chain, we must use a mapping from the cylinder to
the infinite plane to understand the correlation functions. In our simulations, 
we use ground state projector quantum Monte Carlo (QMC) simulations, which means that
we project on a trial state with a large number of Hamiltonians, effectively
for a long imaginary time. This implies that we are using a cylinder whose
circumference is system size and length is infinite for all practical
purposes. The system can then be mapped to the infinite plane thourgh the 
transformations $\tau\to r$ and $x\to \theta$ \cite{Cardy}
which results in two spatially separated points
on the periodic chain having a conformal distance between them of
\begin{equation}\label{ConfDist}
|z_{ij}|=L \sin\Big(\pi\frac{x}{L}\Big),
\end{equation}
where $x$ is the separation on the ring. This substitution into the correlation
functions on the plane tells us what we should expect on the cylinder. 
In some cases, the correlation function on the plane may have different
dependence on $z$ and $\bar{z}$ and may not be expressible in $|z|$, but
the mapping to the cylinder correlation functions reverses this
\cite{Fran,Cardy} and they only depend on the conformal distance 
given by Eq.\eqref{ConfDist}, which must be the case as they are real.
For the case of the two point function, this has already been checked for
the spin-1/2 Heisenberg chain and the TFIM chain \cite{me2}.

\begin{figure}[t]
\includegraphics[width=\hsize]{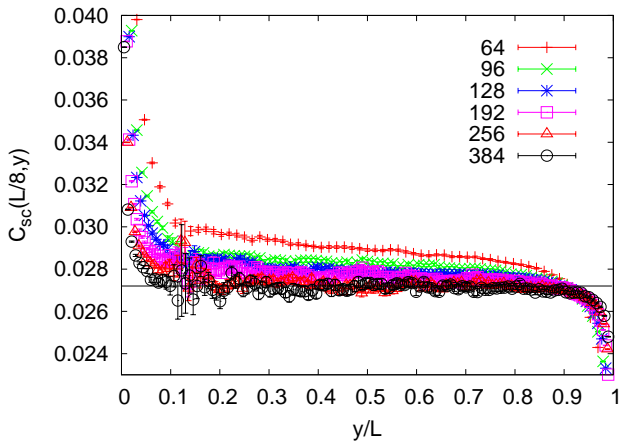}
\caption{Scaled three-point function for the critical periodic TFIM
chain as defined in Eq.~\eqref{TF3pts} flows to a constant for 
$x=L/8$. The constant line is a guide to the eye and the same value for the
constant is used in Fig.~\ref{Fig2}.}
\label{Fig1}
\end{figure}

\begin{figure}[t]
\includegraphics[width=\hsize]{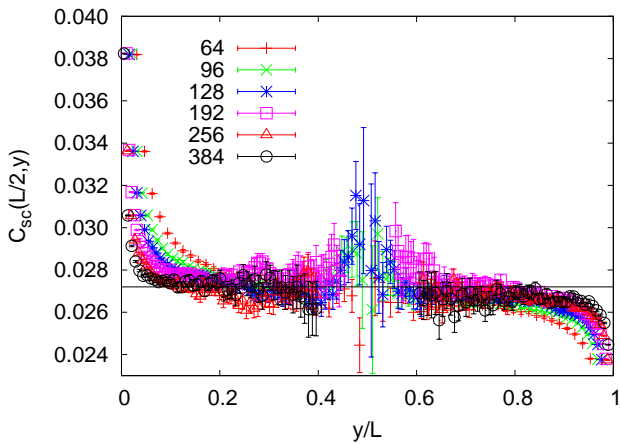}
\caption{Scaled three-point function for the critical TFIM
chain as defined in Eq.~\eqref{TF3pts}  flows to a constant for $x=L/2$.
We only exclude data around $y/L=0.5$ for large sizes as the correlation 
function vanishes quickly as $x\sim y\sim L/2$ and this means the numerical 
signal is very weak.}
\label{Fig2}
\end{figure}

Here we will check that the functional form of the three point function 
matches the CFT prediction for the TFIM as we will be using three point
functions in the next section to point out features of the extended symmetry.
The CFT of the TFIM is made up of three primary operators, namely the
identity $I$, the spin $\sigma$ and the ferromagnetic part of the energy
density $\epsilon$, with scaling dimensions 
0, 1/8 and 1 respectively \cite{241,247}.
With these operators, the non-vanishing three point function with the
smallest scaling dimension
is $\langle \epsilon\sigma\sigma\rangle$. We must also note here
that the CFT only tells us the behavior of the connected three point function
and due to this, we compare the CFT expectation to $\langle \epsilon\sigma
\sigma\rangle_c=\langle \epsilon\sigma\sigma\rangle-\langle \epsilon\rangle
\langle \sigma\sigma\rangle$. We will not carry the subscript $c$ for
connected correlation functions as they make the symbolic
expressions cumbersome.

\begin{figure}[t]
\includegraphics[width=\hsize]{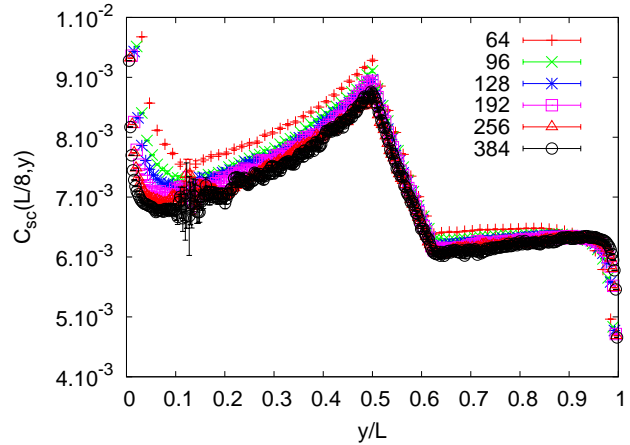}
\caption{Three-point function for the critical TFIM
periodic chain scaled using the lattice distance rather than the conformal
distance as shown in Eq.~\eqref{TF3ptsr}. We see strong disagreement 
with Eq.~\eqref{TF3pt} for $x=L/8$.}
\label{Fig3}
\end{figure}

Using the conformal distance and 
Eq.~\eqref{Corrs2}, the three point correlation function on the ring should be
\begin{equation}\label{TF3pt}
\langle \epsilon_0\sigma_x\sigma_y\rangle\sim
\frac{\big[L\sin\big(\pi\frac{|y-x|}{L}\big)\big]^\frac{3}{4}}
{\big[L\sin\big(\pi\frac{x}{L}\big)\big]\big
[L\sin\big(\pi\frac{y}{L}\big)\big]}.
\end{equation}
To compare numerical data to this expression, we define a scaled correlation
function $C_{sc}(x,y)$ which is the
raw correlation function multiplied by the inverse of the expected form
as shown below:
\begin{equation}\label{TF3pts}
C_{sc}(x,y)\sim
\langle \epsilon_0\sigma_x\sigma_y\rangle\times
\frac{\big[L\sin\big(\pi\frac{x}{L}\big)\big]\big
[L\sin\big(\pi\frac{y}{L}\big)\big]}
{\big[L\sin\big(\pi\frac{|y-x|}{L}\big)\big]^\frac{3}{4}}
\end{equation}
If the expression matches, we should expect the scaled correlation function
to be a constant as a function of $x$ and $y$. In Figs.~\ref{Fig1} and 
~\ref{Fig2}, we plot the scaled
correlation function for two different values of $x$ and the whole range of
$y$ and see that for large sizes, we get a constant and the only deviations
occur when two out of the three operators get close, where the coarse
grained description does not hold anymore. It is important to note here that
we must use the conformal distances when predicting the functional form
due to the cylinder to plane conformal transformation that we have used.
We can also define the scaled correlation function using just the
lattice distances instead of the conformal distance as 

\begin{equation}\label{TF3ptsr}
C_{sc}(x,y)\sim
\langle \epsilon_0\sigma_x\sigma_y\rangle\times
\frac{[s(0,x)][s(0,y)]}
{[s(x,y)]^\frac{3}{4}}
\end{equation}
where $s(a,b)$ is the shortest distance between $a$ and $b$ along the ring.
This way of defining the scaled correlator results
in a disagreement with the expected constant form for the scaled correlation
function, as can be seen in Fig.~\ref{Fig3}, although the curves for
different sizes still show data collapse as the correct scaling dimension
is being used even in this correlator.
This manner of using scaled
correlation functions has already been used in previous work \cite{me2} for
the two point spin correlation function and will also be used in 
the next section to show agreement to the predictions of Sec.~\ref{sec2}.

\section{\label{sec4}Lattice Correlation Function Numerics}

In Sec.~\ref{sec2}, we have seen that some of the correlation functions of
continuum operators vanish and for the ones that do not, we can predict
their functional forms based on the CFT constraints presented in 
Sec.~\ref{sec3}. The primary operators of the $k=1$ WZW model are 
$\bm{J_L},\bm{J_R}$ and $\bm{g}$ with scaling dimensions ($h,\bar{h}$) given
by (1,0),(0,1) and (1/4,1/4) respectively. Using these dimensions,
we can infer that the correlation functions on the periodic chain must have
the following forms,
\begin{equation}\label{ggx}
\langle \mathrm{Tr}[\bm{g}(0)\bm{g^{\dagger}}(x)]\rangle \sim
\frac{1}{L \sin(\pi\frac{x}{L})},
\end{equation}
\begin{equation}\label{JlJlx}
\langle \mathrm{Tr}[\bm{J}_L(0)\bm{J}_L(x)]\rangle \sim
\frac{1}{\big[L \sin(\pi\frac{x}{L})\big]^2},
\end{equation}
\begin{equation}\label{Jlggx}
\langle \mathrm{Tr}[\bm{J}_L(0)\bm{g}(x)\bm{g^{\dagger}}(y)]\rangle \sim
\frac{1}{\big[L\sin(\pi\frac{x}{L})\big]\big[L\sin(\pi\frac{y}{L})\big]}
\end{equation}
and the same for $L\to R$.
We can now use these expressions to understand the lattice correlation
functions by writing the lattice operators in terms of their continuum
versions. Inspired by the analysis leading upto Eq.~\eqref{Latt1}, the spin 
and dimer lattice operators have been postulated \cite{A,Tsvelik} to be

\begin{figure}[t]
\includegraphics[width=\hsize]{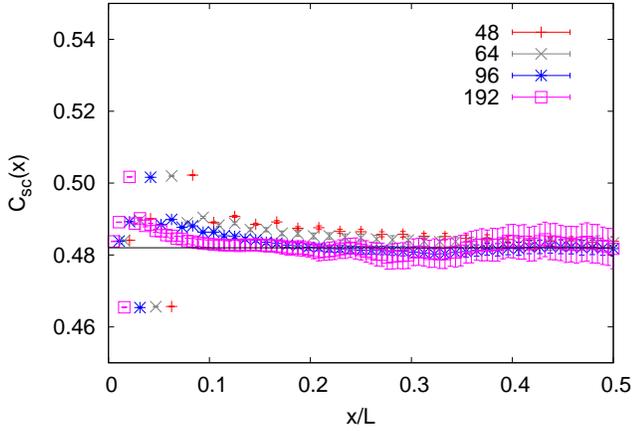}
\caption{Scaled dimer two-point function for the critical $JQ_2$
periodic chain as defined in Eq.~\eqref{DDs} flows to a constant with 
increasing size. Horizontal axis only
extends to 0.5 as two-point functions are symmetric about $y=L/2$.}
\label{Fig4}
\end{figure}

\begin{equation}
\mathbb{S}^i_n \sim \alpha(J_L^i+J_R^i)+(-1)^n\beta S^i,
\end{equation}
\begin{equation}
\mathbb{D}_n=\vec{S}_n\cdot\vec{S}_{n+1}\sim D_0+(-1)^n\gamma D,
\end{equation}
where $\alpha,\beta,\gamma$ are UV sensitive prefactors and $D_0$ is a
constant shift of the lattice dimer operator which must be subtracted out
when calculating connected correlation functions. 
\begin{figure}[t]
\includegraphics[width=\hsize]{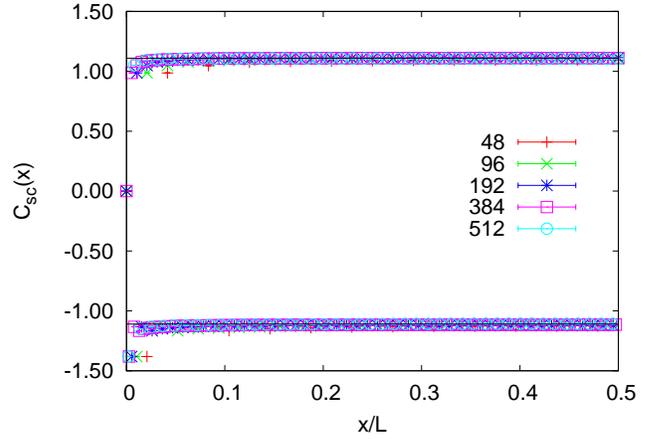}
\caption{Spin two-point function for the critical $JQ_2$
periodic chain scaled with the first term of Eq.~\eqref{SS} 
as shown in Eq.~\eqref{SSs} and we see
agreement with the expected form for large enough sizes. The two branches
capture the relative sign between odd and even sites.}
\label{Fig5}
\end{figure}

\begin{figure}[t]
\includegraphics[width=\hsize]{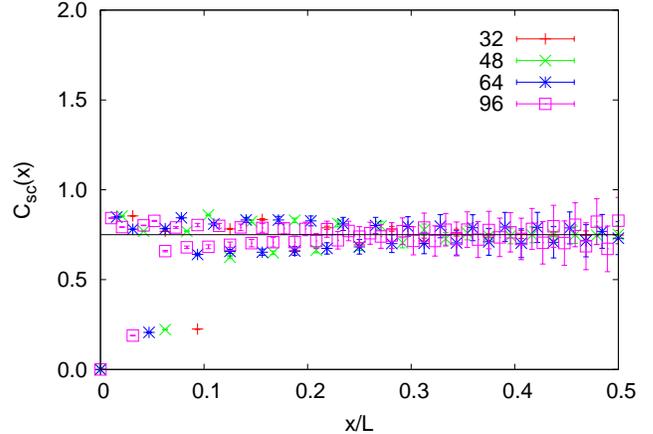}
\caption{Spin two-point function for the critical $JQ_2$
periodic chain with the first term of Eq.~\eqref{SS} subtracted out 
and scaled with the second term as shown in Eq.~\eqref{SS2s}. We see
agreement with the expected form.}
\label{Fig6}
\end{figure}

Using this equivalence
between the lattice and continuum operators, we can construct the lattice
correlation functions that we are going to use to be
\begin{equation}
\langle \mathbb{D}\mathbb{D}\rangle\sim
(-1)^n\langle Tr[\bm{g}\bm{g^{\dagger}}]\rangle+...,
\end{equation}

\begin{equation}
\langle \mathbb{D}\mathbb{D}\mathbb{D}\rangle\sim 0+...,
\end{equation}

\begin{equation}
\langle\mathbb{\vec{S}}\cdot\mathbb{\vec{S}}\rangle\sim
(-1)^n\langle Tr[\bm{g}\bm{g^{\dagger}}]\rangle+
\langle\vec{J}_L\cdot\vec{J}_L\rangle+
\langle\vec{J}_R\cdot\vec{J}_R\rangle+...,\ \ 
\end{equation}

\begin{equation}
\langle\mathbb{\vec{S}}\cdot\mathbb{\vec{S}}\mathbb{D}\rangle\sim
\langle Tr[\bm{J_L}\bm{g}\bm{g^{\dagger}}]\rangle+
\langle Tr[\bm{g}\bm{J_L}\bm{g^{\dagger}}]\rangle+
(L\to R)+...,
\end{equation}
where we have dropped the prefactors as they are UV-controlled parameters
which are not important from the continuum perspective and to keep the
equations from becoming unnecessarily dense. The additional terms ignored
in these equations are lattice corrections which occur due to finite size
lattices and deviations from criticality and our simulations use sizes that
are large enough to justify neglecting these terms. Now we can incorporate the
results of Eqs.~\eqref{ggx},~\eqref{JlJlx} and ~\eqref{Jlggx} to
hypothesize that the full functional forms of the connected lattice correlation 
functions are

\begin{figure}[t]
\includegraphics[width=\hsize]{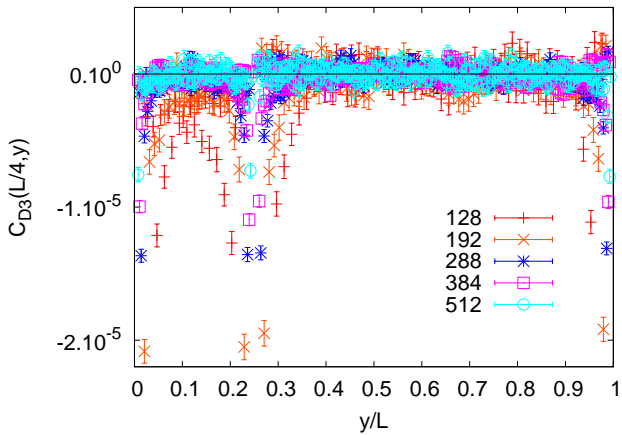}
\caption{Dimer three-point function for the critical $JQ_2$
periodic chain can be seen to vanish for large enough sizes which agrees
with Eq.~\eqref{DDD} for $x=L/4$.}
\label{Fig7}
\end{figure}

\begin{figure}[t]
\includegraphics[width=\hsize]{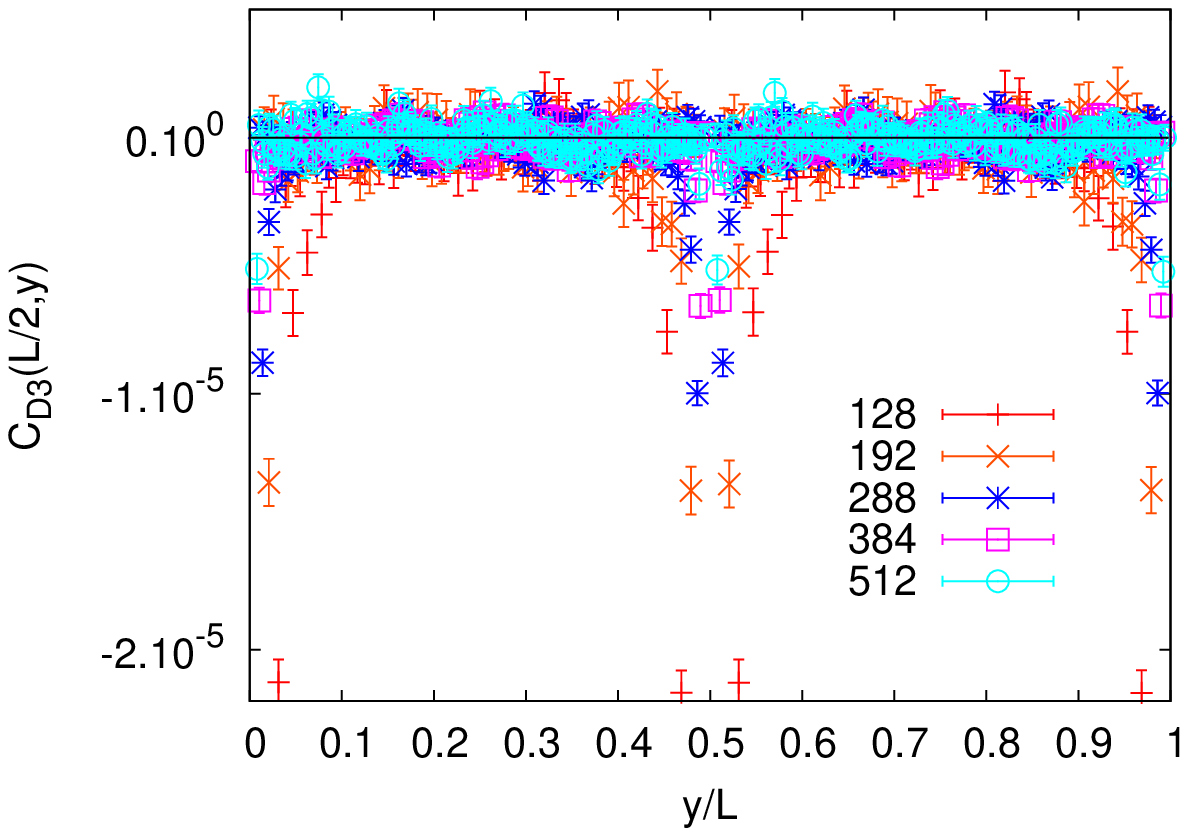}
\caption{Dimer three-point function for the critical $JQ_2$
periodic chain can be seen to vanish for large enough sizes which agrees
with Eq.~\eqref{DDD} for $x=L/2$.}
\label{Fig8}
\end{figure}

\begin{equation}\label{DDD}
\langle \mathbb{D}_0\mathbb{D}_x\mathbb{D}_y\rangle\sim 0+...,
\end{equation}

\begin{equation}\label{DD}
\langle \mathbb{D}_0\mathbb{D}_x\rangle\sim
\frac{(-1)^x}{L \sin(\pi\frac{x}{L})}+...,
\end{equation}

\begin{equation}\label{SS}
\langle\mathbb{\vec{S}}_0\cdot\mathbb{\vec{S}}_x\rangle\sim
\frac{(-1)^x}{L \sin(\pi\frac{x}{L})}+
\frac{1}{\big[L \sin(\pi\frac{x}{L})\big]^2}+...,
\end{equation}

\begin{equation}\label{DSS}
\langle \mathbb{D}_0\mathbb{\vec{S}}_x\cdot\mathbb{\vec{S}}_y\rangle\sim
\frac{1}{L \sin\big[\frac{\pi}{L}(y-x)\big]}
\Bigg[\frac{(-1)^x}{L \sin(\pi\frac{y}{L})}
-\frac{(-1)^y}{L \sin(\pi\frac{x}{L})}\Bigg]+....
\end{equation}

The most striking effects of the extended symmetry are seen in Eqs.~\eqref{DDD}
and ~\eqref{DSS} where the vanishing of the three point function of $\bm{g}$
ensures that there is no term with scaling dimension 3/2 (three times scaling
dimension of $\bm{g}$) in either of these equations.
In this case, if we were to use the three point function's scaling form
to infer the scaling dimensions of the operators (which imply the scaling
dimensions should sum to 2), we would run into errors
as we would be unable to make it consistent with the two point functions
(which imply the scaling dimensions should sum to 3/2).
To see proof of this numerically, we again calculate scaled correlation 
functions for these expressions, except for $\langle \mathbb{D}_0\mathbb{D}_x
\mathbb{D}_y\rangle$, which is expected to be zero. If the numerics agree, we
should expect to see that the scaled functions are constants with respect to
$x$ and $y$, as seen in Sec.~\ref{sec3} for the TFIM.

\begin{figure}[t]
\includegraphics[width=\hsize]{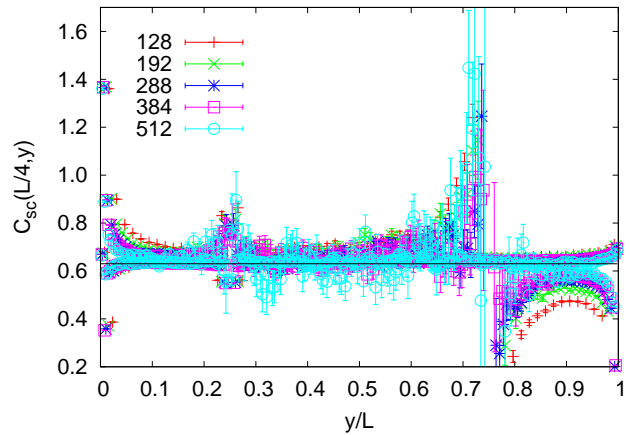}
\caption{Scaled dimer-spin-spin three-point function for the critical $JQ_2$
periodic chain as defined in Eq.~\eqref{DSSs} flows to a constant for $x=L/4$. 
The divergence at 
$y=3L/4$ is caused due to the vanishing of the correlation function, which
implies that the unscaled numerical signal is weak.}
\label{Fig9}
\end{figure}

\begin{figure}[t]
\includegraphics[width=\hsize]{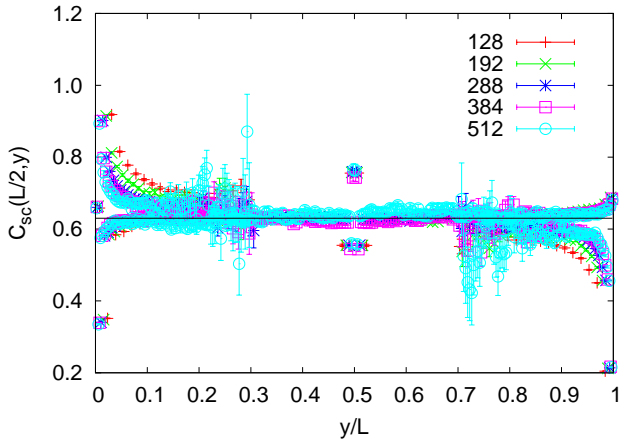}
\caption{Scaled dimer-spin-spin three-point function for the critical $JQ_2$
periodic chain as defined in Eq.~\eqref{DSSs} flows to a constant 
for $x=L/2$. Data for even values 
of $y$ in the range $y/L\in(0.3,0.7)$ are excluded as 
$\langle \mathbb{D}_0\mathbb{\vec{S}}_x\cdot\mathbb{\vec{S}}_y\rangle$ 
tends to a $0/0$ form in that range and thus the scaled correlation function 
is very noisy. }
\label{Fig10}
\end{figure}

The continuum description of the Heisenberg model ground state has
marginal operators which lead to log corrections to correlation functions.
We shall use the $JQ_2$ chain which is the Heisenberg model with a four spin term 
that enforces dimer order when strong and tunes out the log corrections at the
transition point (where the marginal operator vanishes) into the dimer phase;
\begin{equation}
H=-J\Sigma_i \mathbb{P}_{i,i+1}-Q\Sigma_i \mathbb{P}_{i,i+1} 
\mathbb{P}_{i+2,i+3}
\end{equation}
where $\mathbb{P}_{i,j}=1/4-\mathbb{\vec{S}}_i\cdot\mathbb{\vec{S}}_j$.
This model is an alternative to the more commonly used $J_1$-$J_2$ (first and
second neighbor interacting) Heisenberg chain \cite{eggert96}, with the advantage
that it is amenable to QMC studies without sign problems.

At a critical value of $Q/J$, $Q_c/J \approx 0.84831$ \cite{TanSan,sanyal2011}, 
we would expect to see the correlation functions behave in the predicted forms.
All our simulations of the ground state of the critical $JQ_2$ chain 
are done using a projector QMC method formulated in the valence-bond basis. 
The correlation functions are evaluated using loop estimators on the
transition-graphs created by sampling the states in the valence-bond basis \cite{Beach2006}. 
Fig.~\ref{Fig4} 
illustrates the scaled correlator for $\langle \mathbb{D}_0
\mathbb{D}_x \rangle$, defined using Eq.~\eqref{DD} as 
\begin{equation}\label{DDs}
C_{sc}(x)=
\langle \mathbb{D}_0\mathbb{D}_x\rangle\times
\frac{L \sin(\pi\frac{x}{L})}{(-1)^x},
\end{equation}
and we see that it goes to a constant for fairly small
chain lengths. Fig.~\ref{Fig5} shows the scaled version of the first term 
(scaling dimension of 1) 
of $\langle\mathbb{\vec{S}}_0\cdot\mathbb{\vec{S}}_x\rangle$, again defined
using Eq.~\eqref{SS} as 
\begin{equation}\label{SSs}
C_{sc}(x)=
\langle\mathbb{\vec{S}}_0\cdot\mathbb{\vec{S}}_x\rangle\times
\bigg[L \sin\bigg(\pi\frac{x}{L}\bigg)\bigg],
\end{equation}
which dominates the second term (scaling dimension of 2) and we see that
this flows to a constant (1.11(1)) with increasing size. In Fig.~\ref{Fig6}, 
we subtract out
the first term and present the scaled version of the second term in a scaled
correlation function defined as
\begin{eqnarray}\label{SS2s}
&&C_{sc}(x)= \bigg[\langle\mathbb{\vec{S}}_0\cdot\mathbb{\vec{S}}_x\rangle-
1.11\times\frac{(-1)^x}{L \sin(\pi\frac{x}{L})}\bigg] \\
&&~~~~~~~~~~\times\bigg[L \sin\bigg(\pi\frac{x}{L}\bigg)\bigg], \nonumber
\end{eqnarray}
for which we
cannot go to large sizes due to insufficient data quality, and we see that
it matches our expectations and flows to a constant with increasing size. 
In Figs.~\ref{Fig7} and \ref{Fig8},
we show the agreement of the three point dimer correlation function 
(denoted by $C_{D3}(x,y)$ in both figures) with
Eq.~\eqref{DDD} for two different values of $x$ and the whole range of $y$
values. Only in the case of the three point dimer function, we present the
raw correlation function without scaling as it is expected to be zero and there
is no sense in which we can scale it. 
We show the same for Eq.~\eqref{DSS} through 
Figs.\ref{Fig9} and ~\ref{Fig10} by again defining a scaled correlator 
$C_{sc}(x,y)$ as
\begin{equation}\label{DSSs}
\langle \mathbb{D}_0\mathbb{\vec{S}}_x\cdot\mathbb{\vec{S}}_y\rangle\times
\Bigg[\frac{1}{L \sin\big[\frac{\pi}{L}(y-x)\big]}
\Bigg[\frac{(-1)^x}{L \sin(\pi\frac{y}{L})}
-\frac{(-1)^y}{L \sin(\pi\frac{x}{L})}\Bigg]\Bigg]^{-1}
\end{equation}
and observing that it approaches a constant for large sizes
. With this numerical evidence, we 
conclude that the signatures of the extended symmetry that we expect to see are indeed present 
in the spin-1/2 Heisenberg chain. Finite-size (distance) corrections can be seen clearly
in our numerical data and these should be described by irrelevant operators.

\section{\label{sec5}Conclusion}

We have shown numerical evidence of the effects of the emergent SO(4) $\equiv$ [SU(2)$\times$
SU(2)]$/Z_2$ symmetry in the Heisenberg chain
on the correlation functions of lattice operators. This establishes the
IR emergent symmetry which was theoretically expected from a variety of
arguments. The three point function is discussed here as a useful tool
to understand the structure of the underlying field theory and has been shown
to yield useful information through not only its scaling dimension, but also
its functional form, an example of which is 
$\langle \mathbb{D}_0\mathbb{\vec{S}}_x\cdot\mathbb{\vec{S}}_y\rangle$
whose observed scaling dimension is not directly related to
the leading scaling dimensions of the operators it is made out of.
Here, we have also developed tests of CFT through correlation
functions and these can be used to test suspected extended symmetry in higher
dimensional systems\cite{Nahum,NahumX} and more broadly to test for CFT signatures in
general. 

In higher dimensions, an open question is what system geometry is best suited
for investigating the conformal symmetry explicitly in equal-time correlation functions.
In 1D, the ring geometry (infinite imaginary time, i.e., the ground state),  the conformal distance 
naturally emerges in this case, but it does not have a direct generalization to 2D or 3D. The functional 
form of the correlation functions when expressed using the conformal distance in 1D provides perhaps 
the most striking evidence of conformal invariance in finite systems, and such a concept in
higher dimensions would be very useful for lattice calculations.


\section{Acknowledgements}

We would like to thank Cenke Xu for useful discussions. AWS was supported by the
NSF under Grant No.~DMR-1710170 and by A Simons Investigator Grant. 
The computational 
work was performed using the Shared Computing Cluster administered by
Boston University's Research Computing Services.

\bibliography{ref}

\end{document}